# A Structural Phase Transition in $Ca_3Co_4O_9$ Associated with Enhanced High Temperature Thermoelectric Properties

Tao Wu[1], Trevor A. Tyson[1], Haiyan Chen[1], Jianming Bai[2], Hsin Wang[3] and Cherno Jaye[4]
[1]Department of Physics, New Jersey Institute of Technology, Newark, NJ 07102
[2]Materials Science and Engineering, University of Tennessee, Knoxville, TN 37996
[3]Materials Science and Technology Division, Oak Ridge National Laboratory, Oak Ridge, TN 37831
[4] Materials Science and Engineering Laboratory, National Institute of Standards and Technology, Gaithersburg, Md 20899

## Abstract

Temperature dependent electrical resistivity, crystal structure and heat capacity measurements reveal a resistivity drop and metal to semiconductor transition corresponding to first order structural phase transition near 400 K in $Ca_3Co_4O_9$. The lattice parameter *c* varies smoothly with increasing temperature, while anomalies in the *a*, $b_1$ and $b_2$ lattice parameters occur at ~ 400 K. Both $Ca_2CoO_3$ and $CoO_2$ layers become distorted above ~ 400 K associated with the metal to semiconductor transport behavior change. Resistivity and heat capacity measurements as a function of temperature under magnetic field indicates low spin contribution to this transition. Reduced resistivity associated with this first order phase transition from metallic to semiconducting behavior enhances the thermoelectric properties at high temperatures and points to the metal to semiconductor transition as a mechanism for improved *ZT* in high temperature thermoelectric oxides.

# I. Introduction

Thermoelectric materials can convert heat to electrical energy directly and vice versa. These materials are of significant importance since they can recycle waste heat (~ 70% of the total primary energy in internal combustion systems) and may have significant impact on cooling[1]. The performance of thermoelectric materials is characterized by $ZT=S^2T\rho^{-1}\kappa^{-1}$, where $ZT$ is dimensionless figure of merit, $S$ is Seebeck coefficient, $\rho$ is resistivity and $\kappa$ is thermal conductivity. The power factor is defined as $P=S^2\rho^{-1}$, with $ZT$ then given by $PT\kappa^{-1}$. The discovery of $NaCoO_2$[2], which has a high power factor, generated much interest in a class of oxide-based thermoelectric materials which are environmental friendly and have high chemical and thermal stability at high temperature. In this class, the cobaltite $[Ca_2CoO_3]_{0.62}[CoO_2]$, known as $Ca_3Co_4O_9$, with incommensurate misfit structure is widely investigated but not well understood. $ZT$ of single crystal $Ca_3Co_4O_9$ was reported to be ~ 1 at 1000 K[3]. Thus it is considered to be a promising *p*-type thermoelectric material[4]. However, the origin of the high value of $ZT$ at high temperatures is not understood. Previous investigations explored the structure of various forms (powders, crystals and thin films) and doping dependence and the impact of these properties on the thermoelectric figure of merit, and *et al*[5-14]. However, they do not provide a detailed mechanism for the origin of the high $ZT$ values at high temperature.

We note that unlike standard thermoelectrics such as PbTe or $Bi_2Te_3$, $Ca_3Co_4O_9$ is a complex incommensurate monoclinic misfit structure (see Fig. 1) with the superspace group *C2/m(0b0)s0*. It consists of two interpenetrating subsystems of a triple-layered NaCl (rocksalt)-type $Ca_2CoO_3$ block (Subsystem 1) and a $CdI_2$-type hexagonal $CoO_2$ layer (Subsystem 2)[5, 9, 15-17]. The subsystems stack along the $c$ axis and share the same $a$, $c$ and $\beta$ lattice parameters: $a$=4.8270(5) Å, $c$=10.8300(2) Å, $\beta$=98.136(1)°. The mismatch of the unit cells along the $b$ axis results in different $b$ lattice parameters for each subsystem: $b_1$=4.5615(2) Å (subsystem 1), and $b_2$=2.8173(1) Å (subsystem 2)[17]. The $CoO_2$ layer is conducting and the insulated $Ca_2CoO_3$ block is regarded as a charge reservoir[18].

The low temperature thermoelectric properties of $Ca_3Co_4O_9$ have attracted much attention since it shows metallic electrical conductivity (~ 63 $\Omega^{-1}cm^{-1}$ at 300 K) while maintaining high Seebeck coefficient (~ 130 μV/K at 300 K) yielding a high power factor. Two electrical resistivity anomalies have been found (see Fig. 2(a)), one near 20 K and one around 400 K[15, 19]. The low temperature transition is associated with a change in magnetic order while the high temperature transition has been suggested to be associated with a Co spin–state. Specifically, the abrupt slope change in $\chi^{-1}$ around 420 K[15] and the thermal hysteresis loop of $\chi^{-1}$ at 380 K[20] was attributed to a spin-state transition of cobalt. Moreover, Muon spin resonance (μSR) measurements indicated the Co ions change the spin-state between intermediate-spin (IS) + low-spin (LS) and intermediate-spin (IS) + high-spin (HS)[21]. Electron energy loss spectroscopy (EELS) suggested that the $Co^{3+}$ ion spin-state transition occurred from LS at room temperature to IS at 500 K[22].

In this work, we focus on the region of resistivity anomaly around 400 K since this impacts the high temperature thermoelectric properties. The electrical transport behavior change and resistivity drop near 400 K contributes greatly to *ZT* enhancement at high temperature because the thermal conductivity remains esentially constant and Seebeck coefficient (Fig. 2) increases monotonically from room temperature to 900 K[3, 23, 24]. Hence it is important to explore the origin of the changes near 400 K to further improve the thermoelectric properties of $Ca_3Co_4O_9$.

It is of great importance to investigate whether a structural transition induces the resistivity change and to determine the atomic level changes associated with the temperature dependence of the $Ca_3Co_4O_9$ resistivity. A first-order phase transition inducing the metal-semiconductor transition of $Ca_3Co_4O_9$ was proposed by differential scanning calorimetry (DSC) measurements[25]. However, no direct structural experiment has been done to substantiate the nature of this phase transition. In this work, temperature dependence of resistivity and Seebeck coefficient of $Ca_3Co_4O_9$ were measured. Hall measurements were performed to determine the carrier concentration. At the same time, both x-ray diffraction (XRD) and heat capacity measurements as a function of temperature were conducted to explore the structural effects on the electrical and thermal transport. We have found that this transition near 400 K alters the nature of the resistivity curve from one with increasing resistivity with temperature (metal like) to a semiconductor like behavior which enhances power factor. The structural origin of the changes near 400 K is also explored in this work. The reduced resistivity and metallic to semiconducting behavior change associated with this first order phase transition enhance the thermoelectric properties at high temperatures. This result gives an example of thermoelectricity enhanced by a metal to insulator (semiconductor) transition as a route new class of thermoelectric materials.

## II. Experiments

Single phase polycrystalline $Ca_3Co_4O_9$ was prepared by solid state reaction. Stoichiometric amounts of high purity $CaCO_3$ and $Co_3O_4$ were ground, mixed thoroughly and calcined at 900 ℃ in air for 24 hours. The mixture was reground and pressed into a pellet. The pellet was heated at 920 ℃ in air for 24 hours. The latter step was repeated and the final pellet was annealed at 750 ℃ in oxygen flow for 36 hours. The resistivity, Seebeck coefficient and heat capacity measurements of $Ca_3Co_4O_9$ up to 400 K were carried out with a Physical Property Measurement System (PPMS, Quantum Design) with magnetic fields up to 8 T. High temperature Seebeck coefficient measurements up to 710 K were conducted on a ZEM-3 at the High Temperature Materials Laboratory (HTML), Oak Ridge National Laboratory. Temperature dependent Hall effect was measured by PPMS as well. Hall coefficient was obtained by linear fitting the hall resistivity vs. magnetic field data and carrier concentration was calculated. High temperature resistivity was measured by standard four-probe method between 300 K and 710 K in air in a tube furnace using a constant current source and digital voltmeter. *In situ* temperature dependent x-ray diffraction (XRD) measurements ($\lambda$ = 0.7750 Å) between 346 K and 450 K were conducted at beamline X14A at the National Synchrotron Light Source (NSLS), Brookhaven National Laboratory. Finely ground $Ca_3Co_4O_9$ powder (< 25 μm) was loaded into a glass capillary with an inner diameter 0.6 mm and was heated by programmed controlled resistive heating. Rietveld refinement was performed on the XRD data by JANA 2006 [26] .

## III. Results and discussion

Fig. 2(a) shows the temperature dependence of resistivity and Seebeck coefficient of $Ca_3Co_4O_9$ from 2 K to 710 K. Transport behavior changes at ~60 K and ~400 K. The resistivity drops rapidly with increasing temperature below 60 K indicating a semiconducting behavior. Above this region the system exhibits metal-like transport up to 380 K as the resistivity generally increases with temperature. A sudden drop of resistivity occurs near 400 K and transport behavior changes again (reduced resistivity with increasing temperature) up to 710 K. These electrical transport results are consistent with the previously published work[15, 19, 23]. The solid black symbols in Fig. 2(a) show the Seebeck coefficient as a function of temperature. It increases within the whole temperature range and the positive value indicates holes are the major carriers in this system. These results are consistent with the work of Ref. 19. Fig. 2(b) shows the temperature dependent power factor. It increases monotonically within the whole temperature range. The value increases rapidly below 150 K and followed with an almost linear increase. However, a jump is found near 400 K in the shaded window. The enhancement of power factor is attributed to the metallic to semiconducting transport transition (see below). We focus on both transport and structural changes near 400 K.

In order to see the details of the thermoelectric properties change near 400 K. Temperature dependence of resistivity and Seebeck coefficient (Fig. 3(a)) and power factor (Fig. 3(b)) from 355 K to 400 K were shown in expanded form. The circle symbols in open and solid are resistivity in warming up and cooling cycles. It increases monotonically with the increasing temperature up to 16.7 mΩ·cm at 385 K while the slope ($d\rho/dT$) decreased gradually. Then the magnitude of the resistivity drops at the transition. The same trend can be found in cooling. However, it shifts towards lower temperature by ~ 5 K. The transition on warming and cooling processes has an associated hysteresis loop indicating the first order nature of the transition. The black diamond symbols show the Seebeck coefficient on warming. It increases with temperature and drops slightly above the transition temperature. The enhancement of the power factor above the transition is derived from the resistivity drop (Fig. 3(b)). Red lines (guides to the eye) give the slope of the power factor and show the significant enhancement in slope above the transition temperature.

To understand the high temperature transport behavior, high temperature resistivity was fit by the equation

$$\sigma = \frac{\sigma_0}{T} exp(\frac{E_a}{k_B T}),$$

where $\sigma$ is conductivity, $\sigma_0$ is constant, $T$ is temperature, $k_B$ is Boltzmann constant and $E_a$ is activation energy[27]. The linear relationship between $ln(\sigma T)$ and $1/T$ is plotted in the *inset* of Fig. 3(a) and the value of $E_a$ is 56.5 meV. It indicates the semiconducting transport at high temperature.

Fig. 4 is the temperature dependent carrier concentration ($n$) of $Ca_3Co_4O_9$ from 380 K to 400 K with 4 K step size. The Hall coefficient ($R_H$) was obtained from the Hall resistivity as a function of magnetic field (0 T to 8 T). The positive value indicates holes are the major carrier in $Ca_3Co_4O_9$. This is consistent with the Seebeck Coefficient result. Carrier concentration was calculated by $R_H=1/ne$. A sharp increase in carrier concentration is seen above 396K. It indicates the drop of resistivity results from the increase of carrier concentration. Investigation of the structure can shed light on the nature of the transition.

XRD patterns of selected temperatures were shown in Fig. 5(a). The bump marked by arrow is the background from the glass capillary. Two small unindexed peaks are labeled by asterisk (also by D. Grebille *et al*[17]) may originate from an additional modulation vector in the $Ca_2CoO_3$ block. Rietveld refinement was conducted on the temperature dependent XRD data. The superspace group *C2/m(0b0)s0* was employed. The thermal parameters for all oxygen atoms (O1, O2 and O3) were constrained to be equal. The weighted profile $R$ factor ($R_{wp}$) is ~ 3.4% for the temperatures below 400 K and it increases to ~ 3.7% above 400 K. It indicates a structural transition around 400 K. However, the same model was applied to all temperatures since the effect is small. Fig. 5(b) shows the profile of Rietveld refinement on $Ca_3Co_4O_9$ at 346 K. Fig. 1 is the crystal structure of $Ca_3Co_4O_9$. The $Ca_2CoO_3$ block (Subsystem 1) and $CoO_2$ layer (Subsystem 2) with the mismatched lattice parameters $b_1$ and $b_2$ are labeled.

Fig. 6(a), (b), (c) and (d) give the temperature dependence of lattice parameters $a$, $c$, $b_1$ and $b_2$ for the temperature range 346 K - 450 K on warming, respectively. A vertical bar of height 0.002 Å is included in each panel to establish a common scale. Both lattice parameters $a$ and $c$ (Fig. 6(a) and (b)) increase monotonically. The slope of $a$ ($da/d$T) is steeper in the temperature range 390 K – 400 K while lattice parameter $c$ increases linearly with the increasing temperature. The lattice parameter $b_1$ (Fig. 6(c)) increases from 346 K to 390 K. Then it drops from 390 K to 400 K and followed by an increasing trend. In Fig. 6(d), $b_2$ parameter has the same trend with $b_1$. It expands with the increasing temperature, while drops in the transition temperature range (390 K - 400 K). $b_2$ has the biggest error bar compared to the other lattice parameters since it is obtained by $b_1/q$ and the error from both $b_1$ and $q$ was considered. Here $q$ is wave vector to describe the mismatch of two different subsystems and it was obtained directly from Rietveld refinement. The change of the lattice parameters indicates a first order phase transition occurs in the temperature window 390 K - 400 K. Moreover, the change in $x$-$y$ plane dominates the structural transition. We now explore the atomic level changes involved.

Fig. 7 shows the temperature dependent Co-O bonds in both subsystems. Co1-O1 bond (Fig. 7(a)) in $Ca_2CoO_3$ block (NaCl type layer) increases with increasing temperature up to 400 K followed by a fast drop corresponding to the structural transition. It then drops gradually from ~ 420 K. The sudden drop of Co1-O1 bond near 400 K is consistent with the sharp increase of the carrier concentration (see Fig. 4) possibly associated with carrier number modification in the $Ca_2CoO_3$ block considered as charge reservoir. Fig. 7(b) and (c) give the Co2-O3 planar and apical bonds in $CoO_6$ octahedron in $CoO_2$ layer. Co2-O3 planar bond remains constant with increasing temperature and suddenly expands above ~ 390K. It then drops and stays at ~ 1.90 Å to high temperature. Co2-O3 apical bond in Fig. 7(c) increases with temperature up to ~ 390K and then it slowly decreases at high temperatures. Fig. 8 shows the detailed crystal structure of $CoO_2$ layer with the atoms in a $CoO_6$ octahedron labeled. It is composed of a $CdI_2$-type triangular Co lattice with edge sharing $CoO_6$ octahedra and it stacks alternatively with the $Ca_2CoO_3$ block. The change of the Co2-O3 bonds in the $CoO_6$ octahedra indicates it becomes more distorted above ~ 390 K. The distortions in this layer will reduce hopping between Co sites via the oxygen atoms.

Fig. 9 shows the temperature dependent Ca-O1 bonds which are in *x-y* plane in $Ca_2CoO_3$ block. Ca-O1(*y*) (Fig 9(a)) along *y*-axis does not change within the whole temperature range. However, Ca-O1(*x-s*) and Ca-O1(*x-l*) bonds along *x*-axis split by more than 0.05 Å above ~ 390 K. This indicates Ca atoms move to off center positions between neighboring oxygen atoms above the transition. Thus this layer becomes distorted above 390 K which is the transition temperature.

The resistivity and structural parameters all display abrupt changes in the same temperature indicate the first order nature of the transition. The drop of the resistivity at ~ 390 K is connected with the discontinuity of the lattice parameters $a$, $b_1$, and $b_2$ as well as the symmetry in both $Ca_2CoO_3$ block and $CoO_2$ layer. The out-of-plane lattice parameter $c$ changes linearly with temperature. It can be understood in terms of dominant in-plane conduction/resistivity of $Ca_3Co_4O_9$ observed by Masset *et al*[15].

Besides the structural effects, the spin contribution was also investigated. Specifically, resistivity and heat capacity measurements as a function of magnetic field were performed. Fig. 10(a) gives the temperature dependent resistivity of $Ca_3Co_4O_9$ within the temperature region 355 K - 400 K in magnetic fields from 0 T to 8 T with a step size of 2 T. The shape of the hysteresis loop does not change with different magnetic fields. However, the area of the hysteresis loop shrinks with the increasing magnetic field. The area of the hysteresis loop in each magnetic field was calculated to obtain the quantitative area change. The *inset* in Fig. 10(a) shows the relative area change of the hysteresis loop ($A(H)/A(0)$). Here $A(H)$ is the hysteresis loop area in the magnetic field $H$ and $A(0)$ is the area without magnetic field. The area reduces at low magnetic field and then shrinks gradually up to 8 T. Saturation occurs near 4 T. The magnetic field dependence of the hysteresis loop indicates there is a spin contribution to the electrical transport (or that the spin and lattice are coupled). Fig. 10(b) and (c) are the temperature dependent heat capacity with magnetic field in warming and cooling cycles, respectively. The square symbols are for zero field and the triangle symbols are at 8 T. The zero field and 8 T data almost overlap in the whole temperature range on both warming and cooling. Within the statistical errors, the heat capacity does not change with the magnetic field. Thus there is no evidence of spin ordering contribution from the heat capacity measurements. In addition since a significant part of the heat capacity is derived from the phonons spectrum it also suggests that the spin-lattice coupling is weak. We note that the measurements of cobalt $L_2$ and $L_3$ x-ray absorption edges which were conducted by us at 380 K and 412 K reveal no change in the average Co spin state. In terms of the magnitude of spins on the Co sites, it is possible that only a small fraction of Co ion change spin. Electrical transport is sensitive to even small spin state changes since the electron take paths of lowest resistance. The change may be below the level of detectability of our heat capacity measurement.

## IV. Summary

In summary, temperature dependent resistivity measurements on $Ca_3Co_4O_9$ revealed a metallic to semiconducting transition at ~ 400 K. XRD measurement as a function of temperature directly proved the first-order structural phase transition and indicates the associated atomic level changes. The spin contribution to this transition was investigated by resistivity and heat capacity measurements under the magnetic field. It is clear that the hysteresis loop of resistivity shrinks with increasing magnetic field. However, no significant difference of heat capacity with different magnetic field was found. It can be understood that only a small spin state change occurs locally for temperatures near 400 K. The results give an example of high temperature thermoelectricity enhanced by a metal to semiconductor transition and points to a new approach for creating materials with high *ZT* values at high temperature in complex oxide systems. It is important to search for metal oxide materials with metal to semiconductor transitions at high temperature.

## V. Acknowledgments

This work is supported by DOE Grant DE-FG02-07ER46402. The Physical Properties Measurements System was acquired under NSF MRI Grant DMR-0923032 (ARRA award). X-ray diffraction and absorption data acquisition was performed at Brookhaven National Laboratory's National Synchrotron Light Source (NSLS) which is funded by the U. S. Department of Energy.

## Figure Captions

**Fig. 1.** The crystal structure of $Ca_3Co_4O_9$ showing the $Ca_2CoO_3$ block (Subsystem 1) and $CoO_2$ layer (Subsystem 2) with incommensurate $b_1$ and $b_2$ axes labeled.

**Fig. 2.** (a) Temperature dependent resistivity and Seebeck coefficient of $Ca_3Co_4O_9$. (b) Temperature dependence of power factor. The shaded region between 355 K and 400 K shows the metallic to semiconducting transport transition and it is expanded in Fig. 3.

**Fig. 3.** (a) Temperature dependent resistivity and Seebeck coefficient of $Ca_3Co_4O_9$ between 355 K and 400 K. The circle symbols display the resistivity on warming (open) and cooling (solid) with a hysteresis loop. Seebeck coefficient is shown on warming. The *inset* in (a) is the high temperature resistivity data and fit the activation energy $E_a$=56.5 meV. (b) Temperature dependence of power factor on warming. Note that transport transition results in a significant increase of power factor (slope) near 400 K. The straight lines in (b) are guides to the eye.

**Fig. 4.** Temperature dependence of carrier concentration ($n$) of $Ca_3Co_4O_9$.

**Fig. 5.** (a) Synchrotron powder XRD patterns of selected temperatures of $Ca_3Co_4O_9$. The bump marked by the arrow is the background from the glass capillary. Two weak unindexed peaks (*) are discussed in the text. (b) Rietveld refinement of XRD on $Ca_3Co_4O_9$ at 346 K. The observed (crosses), calculated (solid line) and difference (bottom line) profiles are shown.

**Fig. 6.** Lattice parameters of $Ca_3Co_4O_9$ as a function of temperature. (a), (b), (c) and (d) correspond to the lattice parameters $a$, $c$, $b_1$ and $b_2$ between 346 K and 450 K, respectively. Note that anomalies of $a$, $b_1$ and $b_2$ are located in the same temperature range as the abrupt change of resistivity. A vertical bar of height 0.002 Å is included in each panel to establish a common scale.

**Fig. 7.** Temperature dependent Co-O bond lengths including (a) Co1-O1 bond in $Ca_2CoO_3$ block (Subsystem 1), (b) Co2-O3 planar bond and (c) Co2-O3 apical bond in $CoO_2$ conducting layer (Subsystem 2). The structure becomes more distorted above ~ 400 K.

**Fig. 8.** $CoO_6$ octahedra in the conducting $CoO_2$ layer with apical and planar oxygen atoms labeled.

**Fig. 9.** Temperature dependence of Ca-O1 bond lengths in *x-y* plane in $Ca_2CoO_3$ block. (a) Ca-O1 (*y*) shows Ca-O1 bond along *y*-axis. (b) Ca-O1 (*x-l*) and Ca-O1 (*x-s*) are the long and short Ca-O1 bonds along *x*-axis, respectively.

**Fig. 10.** (a) Hysteresis loop of resistivity with magnetic fields up to 8 T. The *inset* shows the relative area of the hysteresis loop as a function of magnetic field. Temperature dependent heat capacity with H=0 T and 8 T in warming (b) and cooling (c) processes, respectively.

**Fig. 1. T. Wu *et al*.**

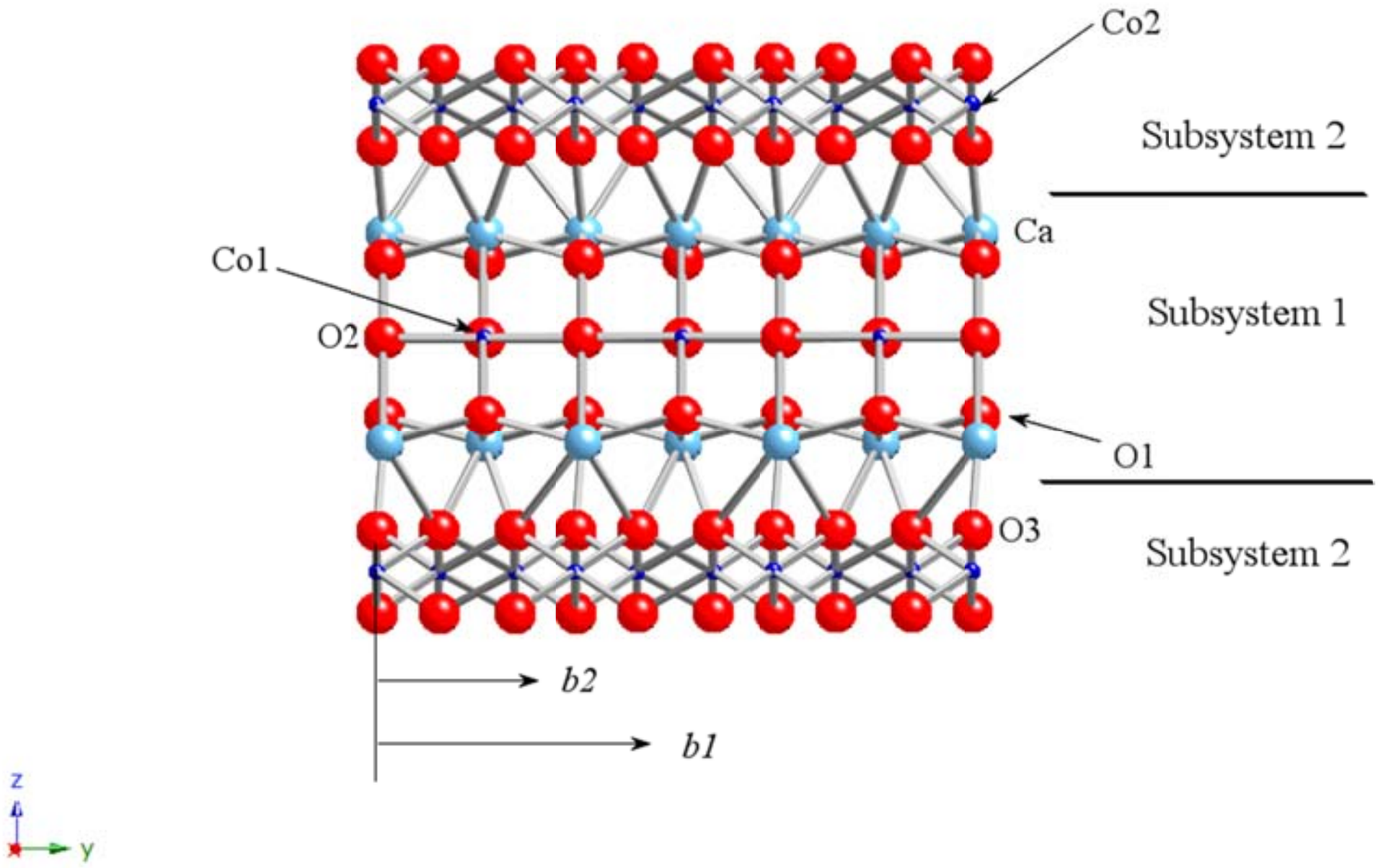

**Fig. 2. T. Wu *et al.***

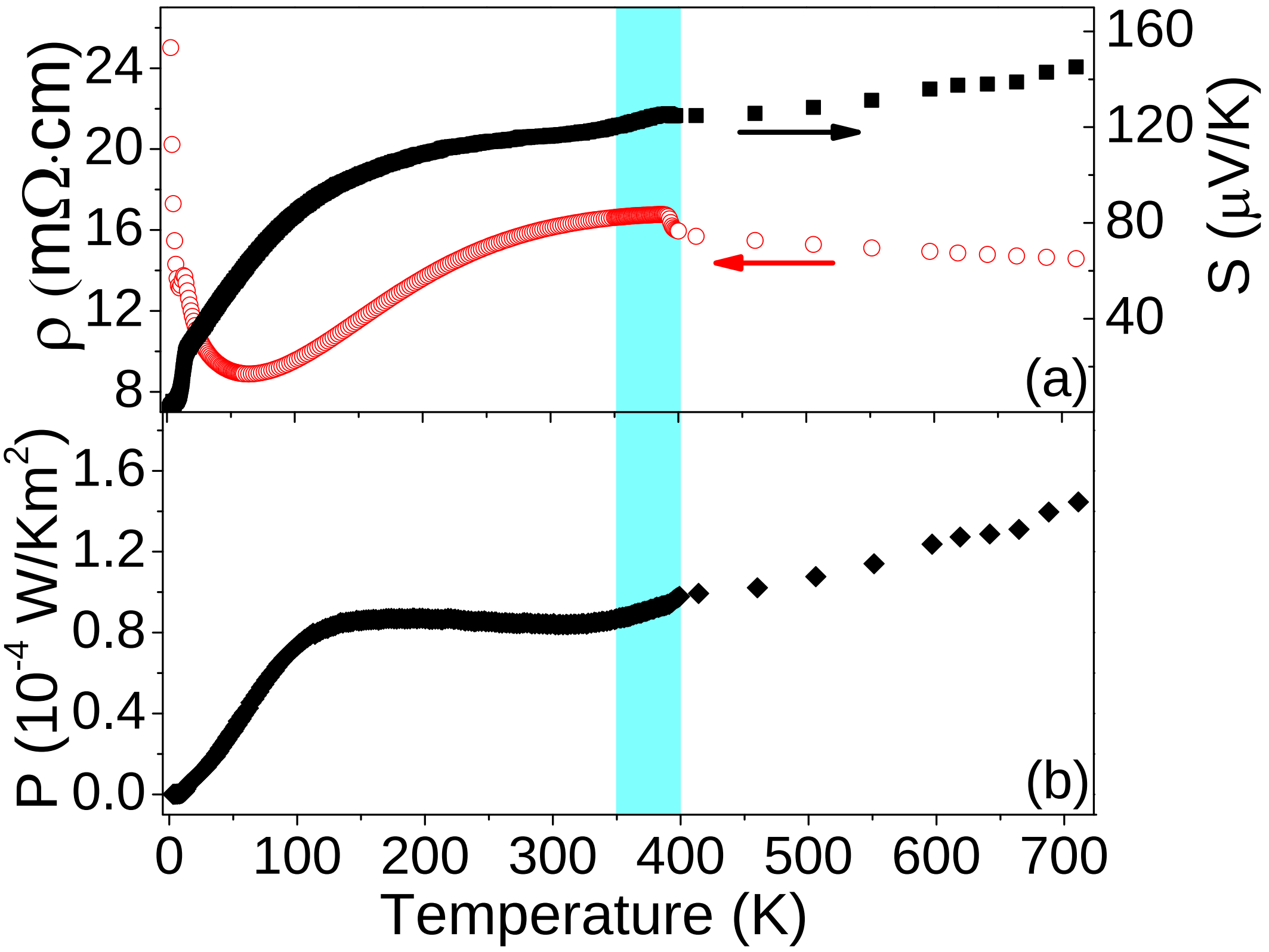

**Fig. 3. T. Wu *et al.***

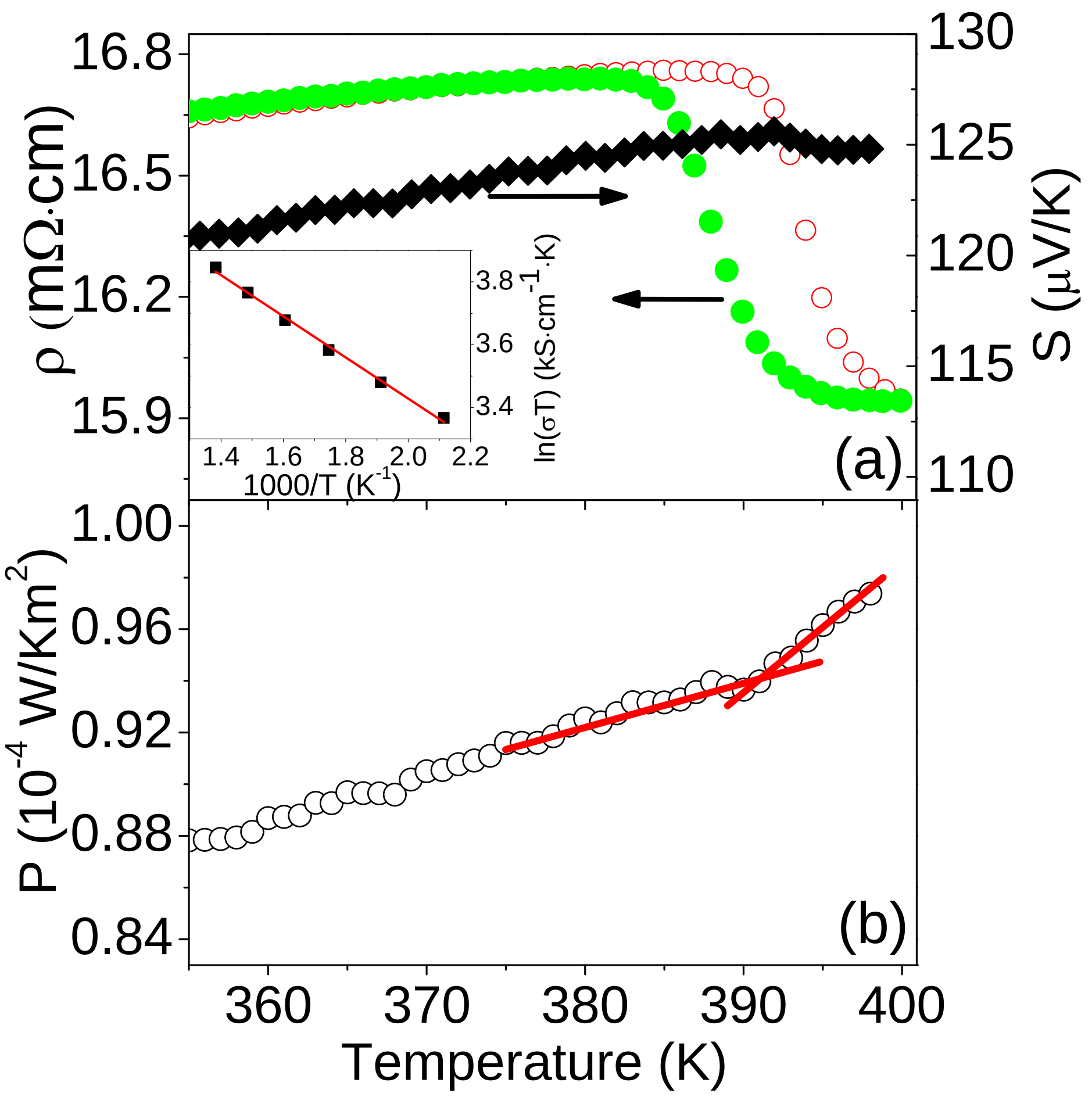

**Fig. 4.** **T. Wu *et al.***

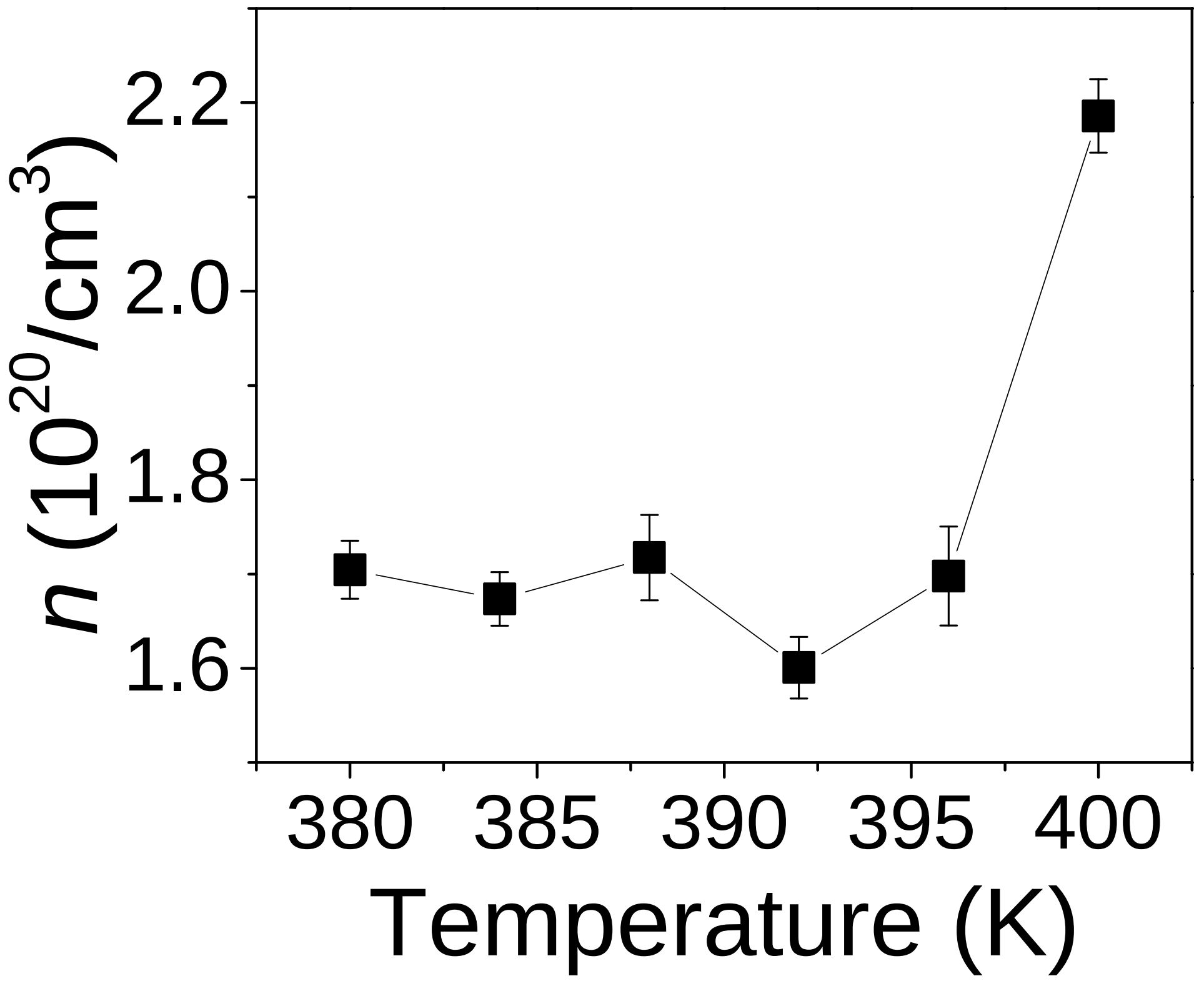

**Fig. 5. T. Wu *et al*.**

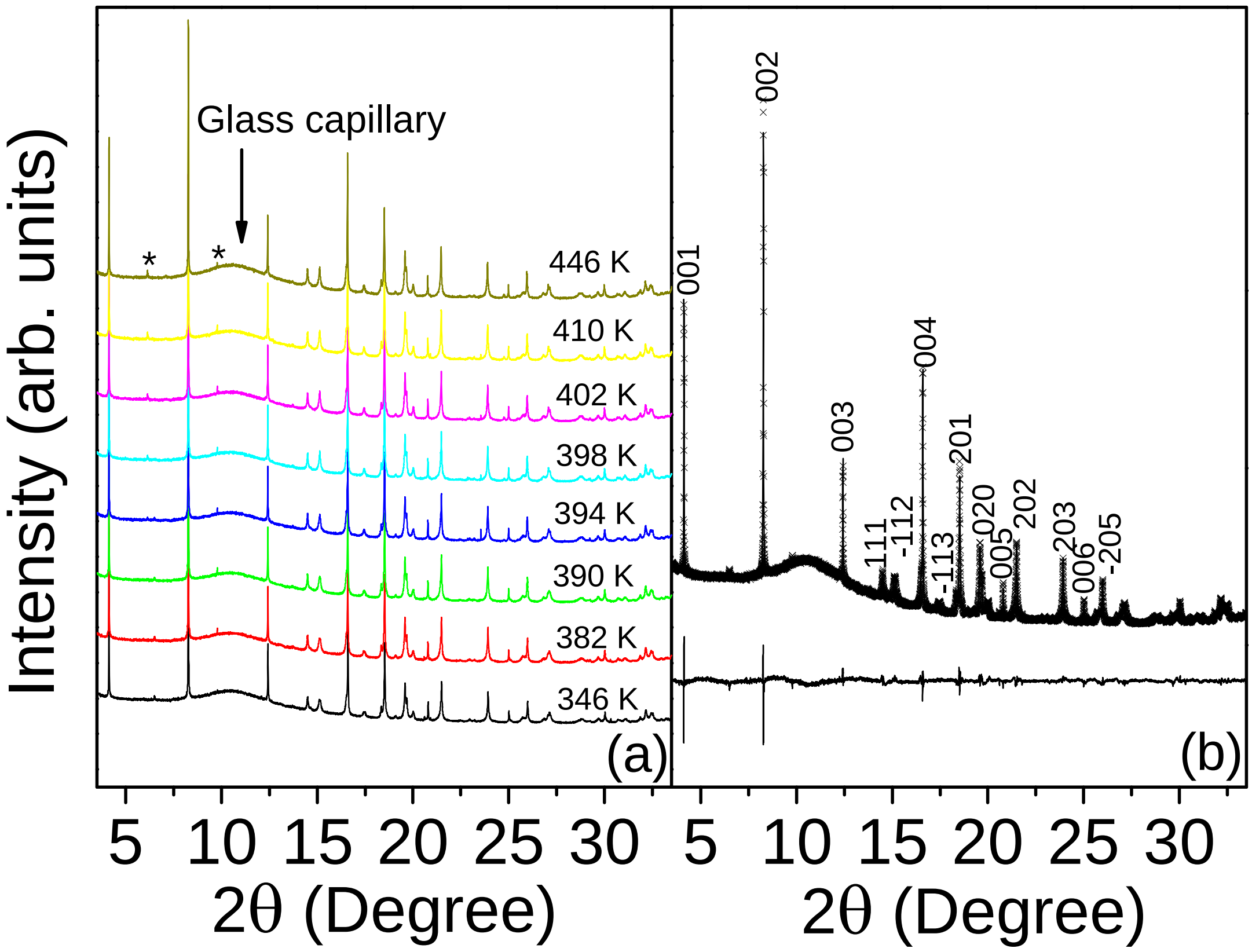

**Fig. 6. T. Wu *et al.***

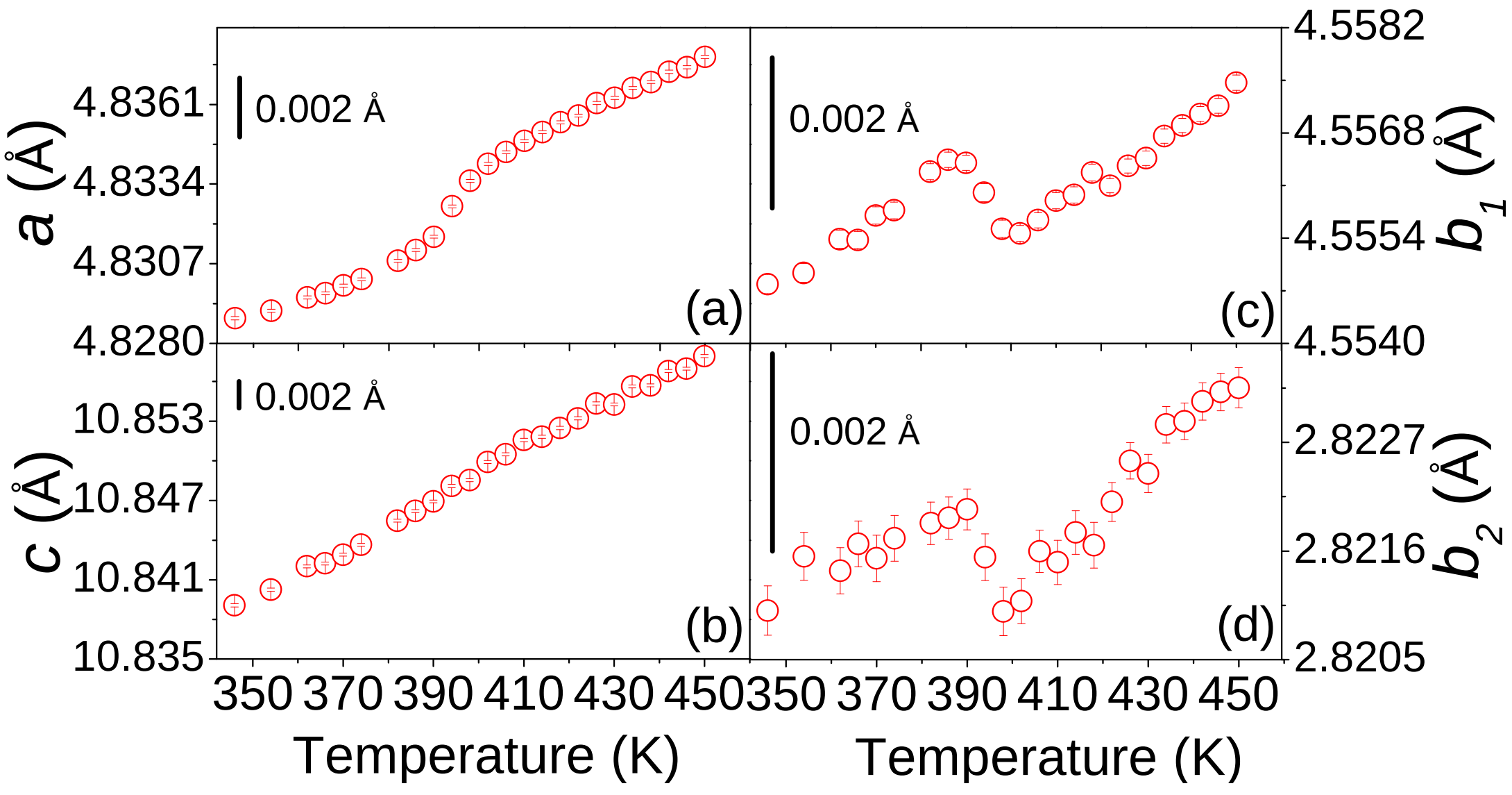

**Fig. 7.  T. Wu *et al.***

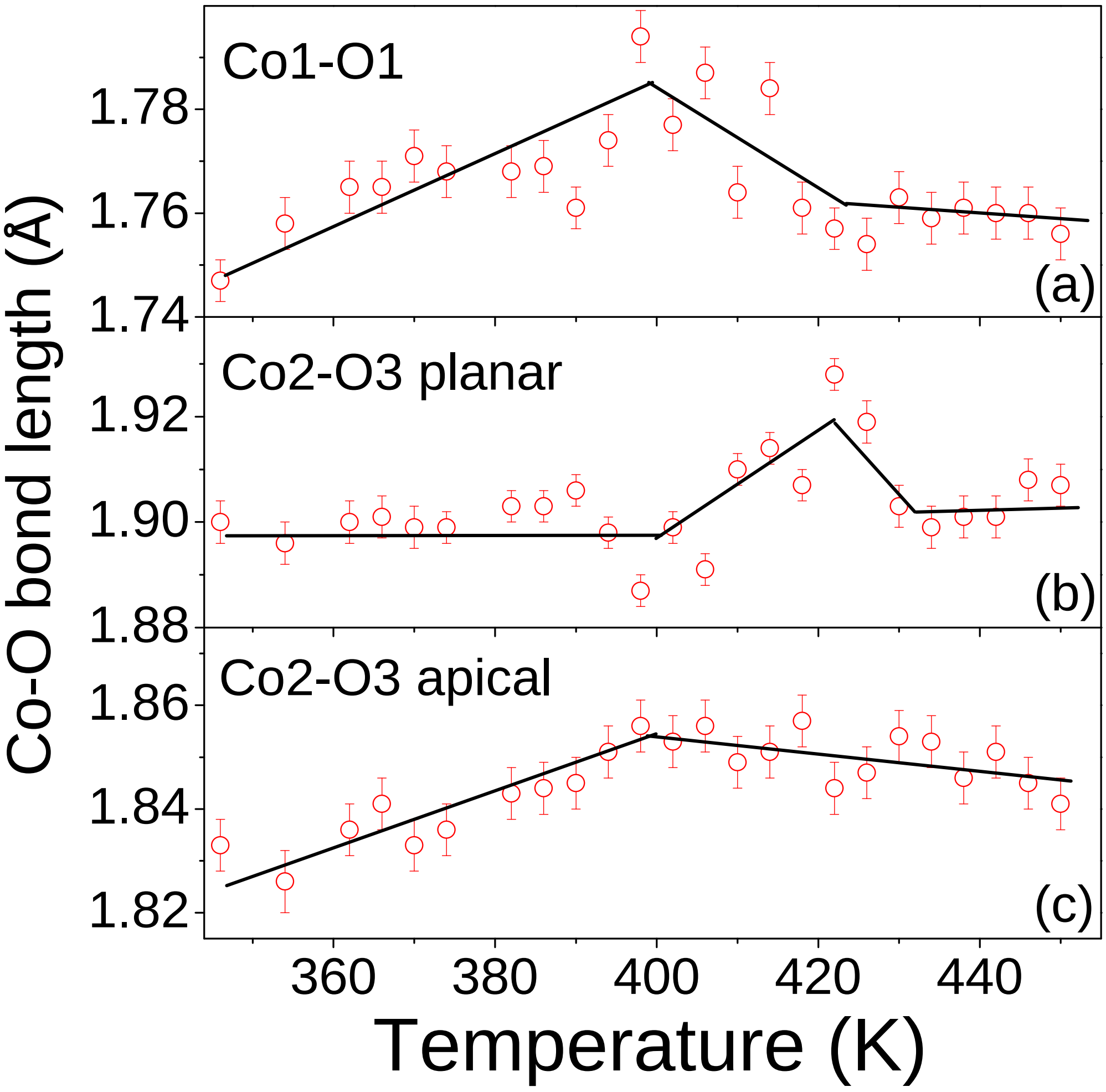

**Fig. 8. T. Wu *et al*.**

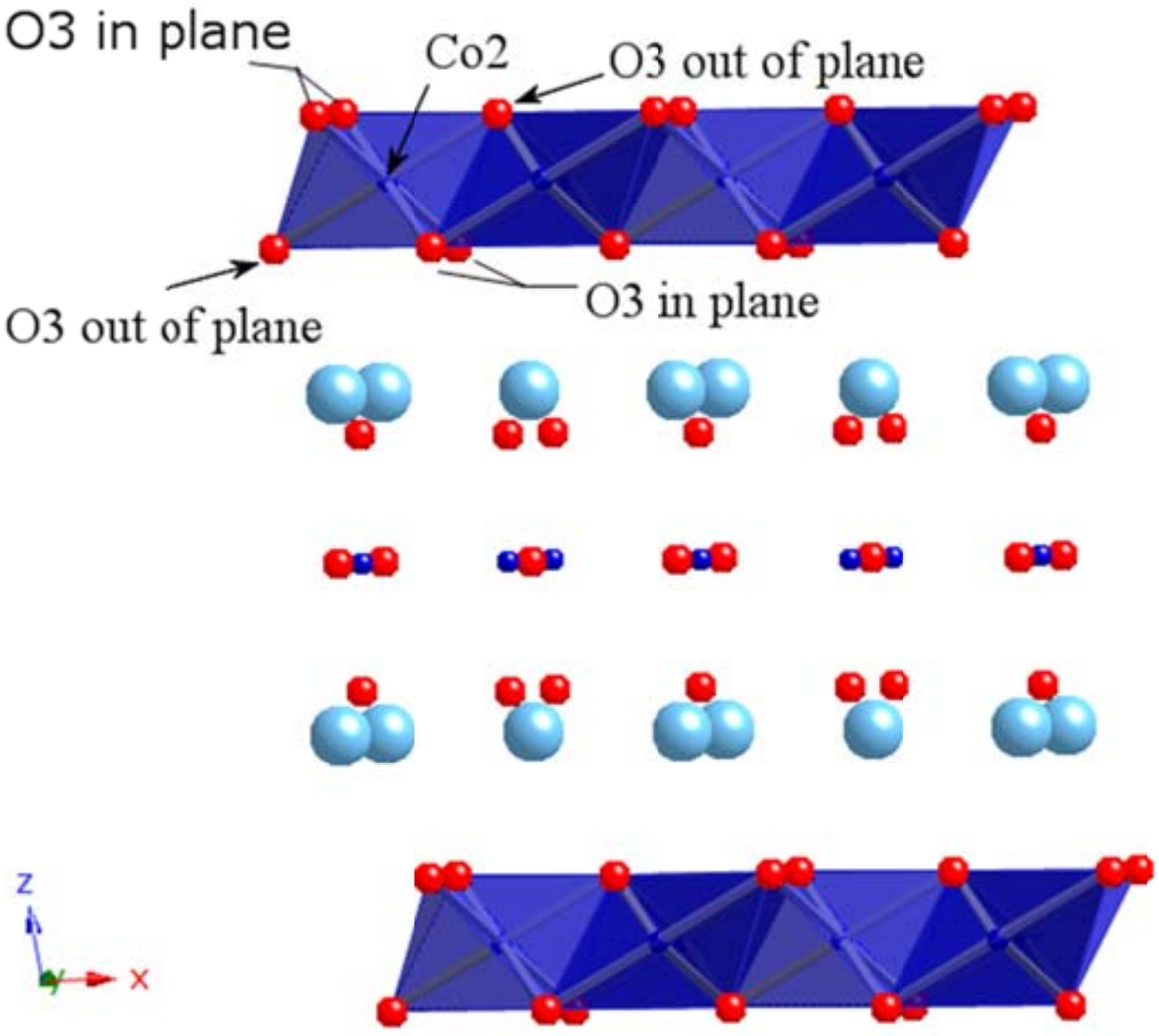

**Fig. 9. T. Wu *et al.***

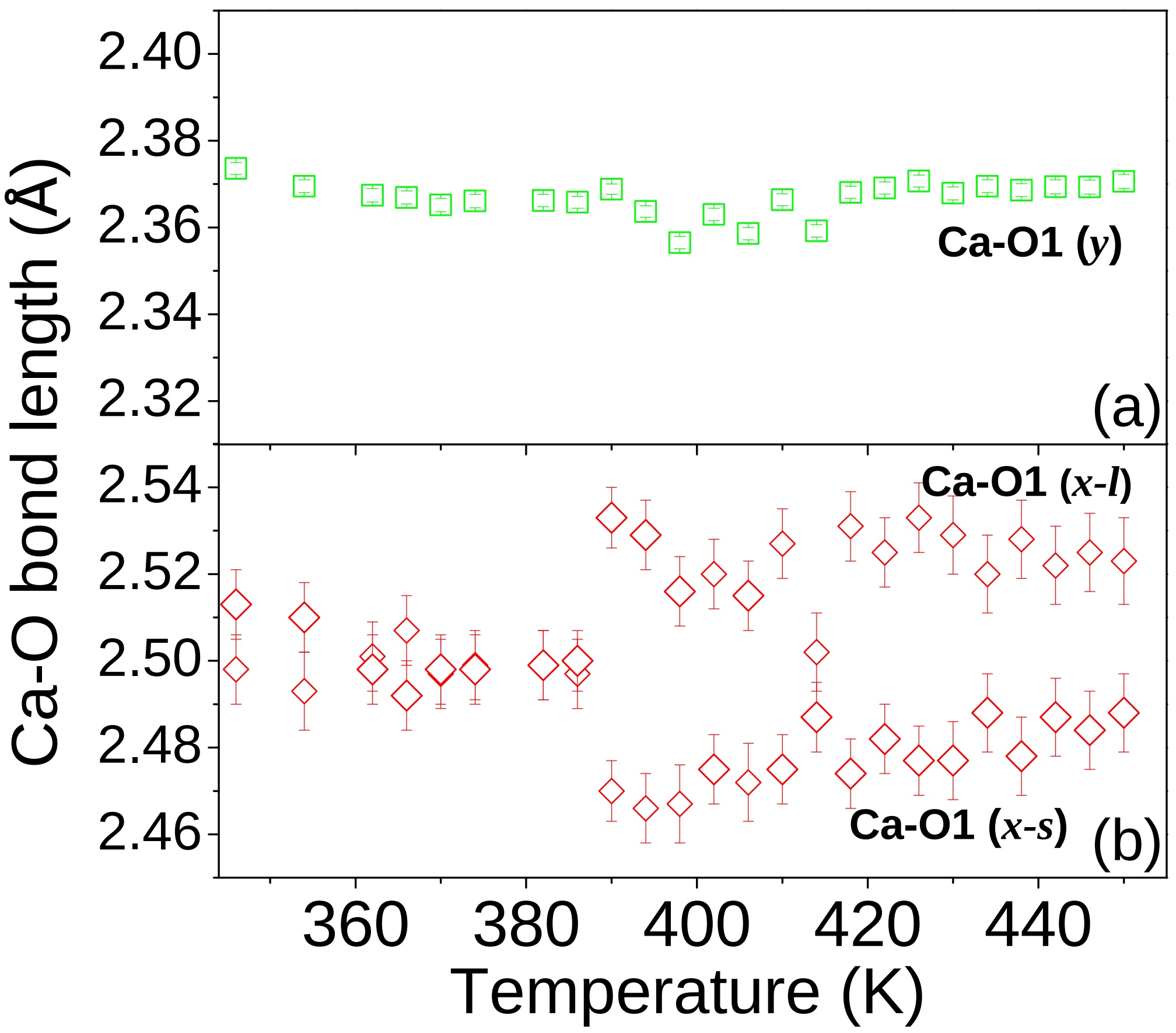

**Fig. 10. T. Wu *et al*.**

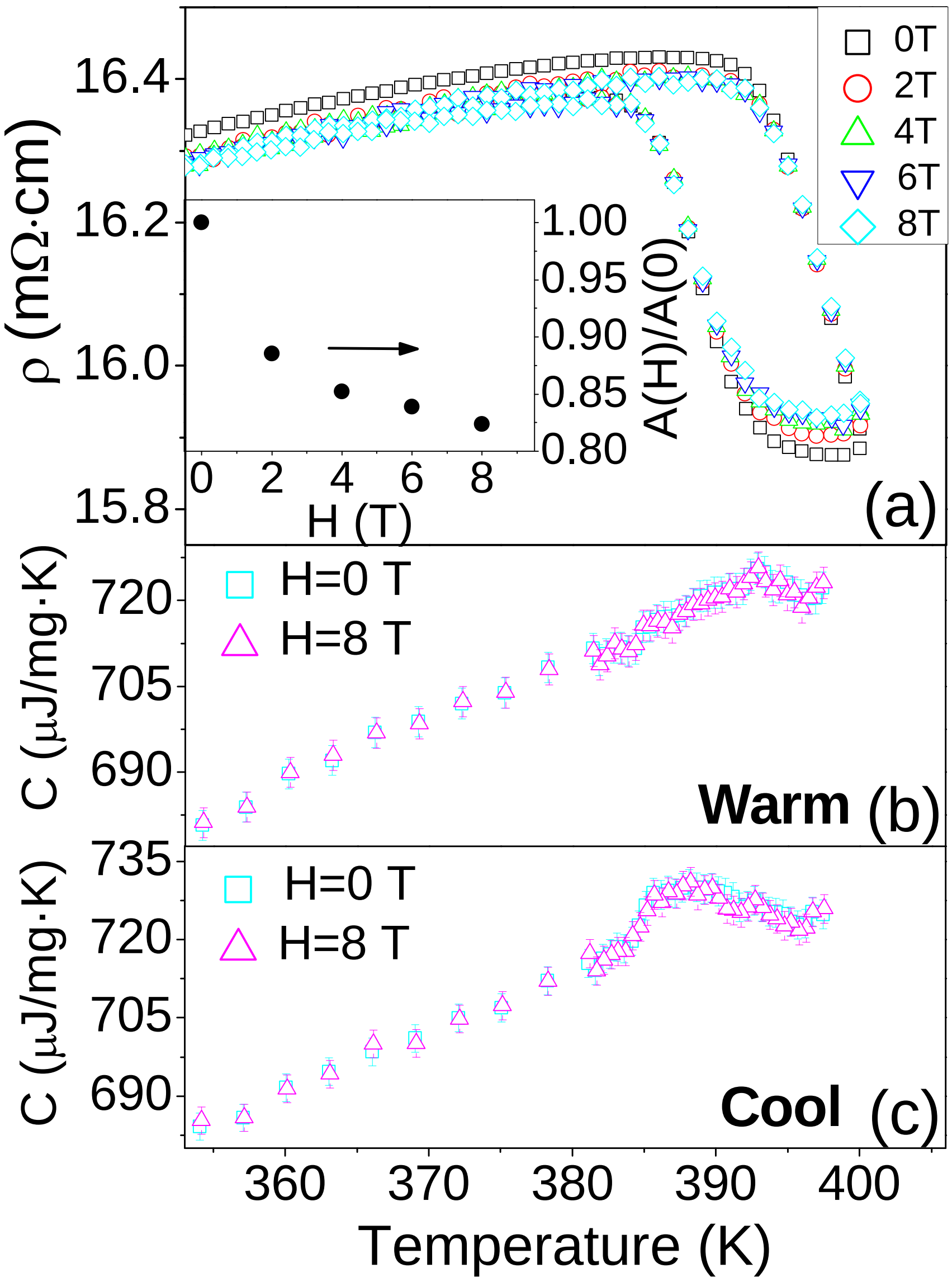